\documentstyle[prl,aps,preprint,fps]{revtex}

\newcommand{\be}{\begin{eqnarray}}
\newcommand{\ee}{\end{eqnarray}}
\newcommand{\lb}[1]{\label{#1}}
\newcommand{\e}{{\bf\varepsilon}}
\newcommand{\et}{\e_{th1}}

\begin{document}
\setcounter{page}{0}
\def\footnoterule{\kern-3pt \hrule width\hsize \kern3pt}
\tighten
\title{Quark states near a threshold\thanks
{This work is supported in part by funds provided by the U.S.
Department of Energy (D.O.E.) under cooperative 
research agreement \#DF-FC02-94ER40818.}}

\author{S.~V.~Bashinsky and R.~L.~Jaffe}

\address{{~}\\Center for Theoretical Physics \\
Laboratory for Nuclear Science \\
and Department of Physics \\
Massachusetts Institute of Technology \\
Cambridge, Massachusetts 02139 \\
{~}}

\date{MIT-CTP-2572,~hep-ph/9610395. {~~~~~} September 1996}
\maketitle

\thispagestyle{empty}

\begin{abstract}

We reduce the problem of many-channel hadron scattering at
nonrelativistic energies to calculations on the scale of a few fermis.
Having thus disentangled kinematics from interior quark dynamics, we
study their interplay when a quark state occurs near a hadronic
threshold. Characteristic parameters, such as the observed peak
width, the decay width, and the shape of a cross-section itself are
highly affected by the threshold.  A general pole-form expression for the
$S$-matrix in an arbitrary background is given, and the pole structure
of $S$ is examined.  We show that at a hadronic threshold two
poles in $S$ are generally important.  We also classify the $S$-matrix pole
structure considering an example where nonsingular coupled channels
are closed at the threshold.  The framework of our paper is the
$P$-matrix formalism, which is reviewed and extended for use together
with conventional methods of computing quark-gluon dynamics.  Results and
applications are illustrated for the doubly strange two-baryon system,
the detailed analysis of which we postpone till our forthcoming paper.
\end{abstract}

\begin{center}
Submitted to: {\it Nuclear Physics B}
\end{center}

\section{Introduction and Summary}
 
A resonance shape can be dramatically distorted if one of its decay 
channels has a threshold within the resonance width.  A tiny variation 
of coupling strength may lead to a wide spectrum of physical phenomena 
such as a slightly bound or a virtual state, a ``shoulder'', or a resonance.  
All these effects are of kinematic origin.  We will show that the underlying 
quark-gluon dynamics can be isolated and quantitatively estimated in a smooth 
way which is unaffected by such kinematic cataclysms.

There is little doubt that far from threshold singularities narrow and 
dramatic effects in scattering amplitudes are to be identified with 
quasi-stable states of QCD.  Little sophistication is required to connect 
the $\rho(770)$ with $\bar u u {-} \bar d d$ or the $\phi(1020)$ with 
$\bar s s$.  
However, great care must be used when attempting to assign a fundamental QCD 
interpretation to broad effects like most of those seen in meson-meson 
scattering above 1 GeV or to striking effects like the $f_0(980)$ and 
$a_0(980)$ that lie near thresholds (in this case $K\bar K$). 
Identification many objects of great interest --- exotics, hybrids, 
glueballs,  quasi-molecular 
states, {\it etc.\/} --- require us to consistently relate low energy 
scattering to microscopic quark-gluon dynamics.

We study hadron-hadron scattering at small kinetic energy, where 
non-relativistic methods suffice.  This is an old problem, but there is no 
general agreement on how to associate quark-gluon ``states'' with effects 
seen in low energy scattering.
One of the most popular phenomenological tools is the 
$K$-matrix parameterization$^{\cite{Levisetti}}$ and its pole analysis. 
The $K$-matrix emerges naturally in the study of 
dynamics that occurs at distances much smaller than the de Broglie wavelength 
of the scattering system. 
For a single channel close enough to threshold, the conditions for a
$K$-matrix analysis might seem to be met.  However, in the real world
hadron-hadron systems with small relative momentum are often strongly
coupled to other open or closed channels, where the relative momentum is
large in absolute magnitude compared to the intrinsic sizes of
hadrons. In this case results obtained from solving microscopic quark
dynamics must not be directly associated with the $K$-matrix.  We hope to make 
this clear in the course of our paper.  In place of the $K$-matrix, we will 
argue that the $P$-matrix formalism$^{\cite{P-orig}}$  is more suitable for 
this purpose.

This paper consists of two general divisions. Section~2 concerns
with microdymanics on the hadron-size scale and the $P$-matrix formalism.
The following Section~3 deals with observable objects such as 
$S$-matrix and cross-sections. Many of our results are outlined in the
rest of the introduction below.

In Section~2A we review and extend the $P$-matrix formalism.
$P$ is defined, similar to $K$, as an algebraic transform of the 
$S$-matrix but it involves an additional parameter $b$:
\be
P(\e,b)=i{~}\sqrt{k}{~}\frac{e^{ikb}S(\e)e^{ikb}+1}
{e^{ikb}S(\e)e^{ikb}-1}{~}\sqrt{k}{~},
\lb{PS}
\ee
where we consider a multichannel $s$-wave with the total energy $\e$.
If the interaction for $r{>}b$ is absent or simple enough to be 
described with a potential, the $P$-matrix generalizes the logarithmic
derivative  of the wave function at $r{=}b$ (see Section 2A for details). 
Then $P$ is fully determined by the dynamics in the inner domain 
$r{<}b$.

The poles of $P(\e,b)$ play an important role. 
Their positions and residues are shown to be related to the spectrum of the 
hadron-hadron system confined in a spherical cavity with a radius $R$ 
depending on $b$. Such boundary conditions are used in the bag model and 
could be simulated on a lattice. 

We will vary the parameter $b$ changing the size of the cavity $R(b)$ and 
tracing the evolution of the $P$-poles. The smaller $R$, the more simple 
the quark 
dynamics inside the cavity. But when $R$ becomes equal or less than the 
confinement radius, the connection between the spectrum of the physical 
system and $P$-poles gets more and more complicated. In practice one has 
to stop at some $R_0(b_0)$.
It was proposed in Ref.~\cite{P-orig} that at this $R_0$ 
the quark system in the cavity may be treated as a single bag and its 
eigenstates
can be calculated in  perturbative QCD with current quark masses. 
This assumption reflects the idea that the bag interior is a phase built up on 
the perturbative vacuum. Alternatively, it could be a phase in which chiral 
symmetry is spontaneously broken, yielding constituent quarks with 
renormalized couplings and pion-like  excitations. Finally one might also 
attempt to exploit lattice methods.

In Section 2B we illustrate the $P$-matrix calculation taking the bag model
as an example. The issue of flavor symmetry is addressed.
We show that $P_{ij}(\e,b)$ reflects this symmetry provided the cavity is 
sufficiently small. Then the flavor projections of the quark-bag states 
onto a two-particle state determine the corresponding projections 
of the $P$-pole residues. The latter, in turn, yield the partial decay width 
of the states, as it is shown in Section 3.

In Section 3A we explore the relation between the $P$-matrix and the
$S$-matrix, concentrating on $S$ poles, their widths and their channel
couplings. We will find a pole form equation for the 
$S_{ij}(\e)$ and thus obtain the formulas for the position, width, and the 
decay 
amplitudes of observed resonances or virtual states.
All this will be done in the presence of an arbitrary 
phenomenological background $S_0(\e)$, and the unitarity of $S$ will 
be preserved. 

At a threshold the momentum $k(\e)$ becomes singular. 
But unless $b$ is too big, the $P$-matrix does not ``feel'' the threshold 
and has a smooth behavior considered in Secs. 2A and 2B.  This enables us to 
separate cusp effects from inner dynamics (in a way similar to $K$-matrix
analysis). In the Section 3B we discuss  the analytical structure of a
many-channel $S$-matrix when a pole in $P(\e)$ occurs near some threshold.
We find that the $P$-pole
gives rise to {\it two} nearby poles in $S$, and track the movement
of these poles on the many-sheeted energy plane, where they appear as a bound,
quasi-bound states, and a resonance as the coupling strength
decreases. We note the drastic energy  dependence of hadronic shift and width 
at a threshold. One of its important consequences is a substantial difference 
between a quasi-bound state decay width and the corresponding observed 
resonance width (peak width).
    
An application of our methods can be found in a forthcoming
paper$^{\cite{BJII}}$ where we consider the low energy production and 
scattering of two baryons  with the total strangeness minus two. 
We will investigate the possibility 
for the 6-quark $H$-dibaryon$^{\cite{H-orig}}$ to be unstable
with respect to strong decay.

We would like to emphasize that we prefer this formalism
to the $K$-matrix parameterization because:
\begin{itemize}
\item
it is simply connected with dynamical calculations. One can, in principle, 
find the $P$-matrix solving a boundary value problem on a microscopic scale;
\item
$P$ obtained this way does not need correction for the coupling with open 
channels (hadronic shift). 
\end{itemize}
$K$ matrix may also be considered for a many-channel system. 
Nevertheless, the connection between quark dynamics
and $K$-pole structure or symmetries is not straightforward, unless at the 
given energy the wavelength is large ($kb_0{<<}1$) in {\it all} of the coupled 
channels. This never can be realized if some of the strongly coupled channels 
have different threshold energies. 
Even for one-channel $NN$ scattering this holds 
only in $10{~}MeV$ energy interval\footnote{The choice of 
$b_0\simeq10^{-2}MeV^{-1}$
is made according to Ref.\cite{P-orig}; see Sec.~2B for details.}
while 
\begin{itemize}
\item
the $P$-matrix formalism gives simple 
dynamical interpretation over the whole nonrelativistic range ($\sim 10^{3}
{~}MeV$).
\end{itemize}
$P$ also inherits the main advantages of the $K$-matrix:
\begin{itemize}
\item
it provides a parameterization of $S$ supporting its unitarity;
\item
it is insensitive to threshold singularities.
\end{itemize}
The formal definition of $P$ by eq.~(\ref{PS}) resembles the definition of $K$.
In fact,
\be
K(\e)=P^{-1}(\e,b{=}0)
\lb{KP}
\ee
which implies that 
\begin{itemize}
\item
many analytical results in the well investigated $K$-matrix formalism
are directly generalized to $P$-matrix.
\end{itemize}
\section{The P-matrix}

This method of analyzing two-body reactions was proposed by Jaffe and Low 
in 1979 in order to test the spectroscopic predictions of quark models
especially as they relate to exotic ({\it e.g.\/} multi-quark) states. 
The method was initially developed in the context of the bag model, where 
quarks are confined by a scalar vacuum pressure. However, it applies to any 
model in which quark and gluon eigenstates are studied without considering 
their coupling to decay channels.
 First we briefly review their
formalism.  We also present new arguments that  give a further insight into the
connection between low-energy scattering and  quark model speculations. In
the next subsection we quantitatively estimate  the parameters of the 
$P$-matrix from the quark-bag model. 

\subsection{Formalism}

At low kinetic energies hadron-hadron scattering may be described
by nonrelativistic kinematics. Restricting our attention to $S{=}L{=}0$,
we factor out the center-of-mass motion and consider the wave-function of 
the $n$-channel two-hadron system in the relative coordinate $r$.
For a given value of a spatial parameter~$b$, a definite energy~$\e$, 
and $r$ greater than the interaction radius, the most general form of the
wave function is  
\be
\psi_i(r_i)=\sum_{j=1}^n\left\{\cos[k_i(r_i-b)]{~}\delta_{ij} 
+\sin[k_i(r_i-b)]{~}\frac{P_{ij}}{k_i} \right\} A_j, 
\lb{defP}
\ee
where $i=1,{~}...{~},n$ labels the channel and the  $\{A_j\}$ 
are some amplitudes.

The matrix $P_{ij}$ generalizes the logarithmic derivative of $\psi(r)$ 
for the case of many channels. Comparing eq.~(\ref{defP}) to the usual 
$S$-matrix parameterization of the scattering wave function, we find that $
P$ and $S$-matrix are simply related as$^{\cite{P-orig}}$:
\be 
S=e^{-ikb} \frac{{ \frac{1}{\sqrt{k}}} P
 {\frac{1}{\sqrt{k}}} +i}{{\frac{1}{\sqrt{k}}}P
 {\frac{1}{\sqrt{k}}}-i} e^{-ikb}{~}.
\lb{SP}
\ee
Resolving this equation with respect to $P$ one gets eq.~(\ref{PS}).
For a unitary $S$ the matrix $P$ is hermitian. If the interaction is time
reversal invariant, then $P$ is also real. 
$P$ depends on $b$ according to the equation
\be
\frac{\partial P}{\partial b}=-P^2-k^2{~}.
\lb{dP/db}
\ee
  
For the present we treat $b$ as a free parameter. Suppose for a moment 
that the value of~$b$ is large enough so that there is no interaction for 
$r\geq b$. If for the energy $\e=\e_p(b)$ and some choice of the amplitudes 
$A_j$
in eq.~(\ref{defP})
\begin{figure}
\fpsxsize=1.5in
\def\fpsangle{90}
\centerline{\fpsfile{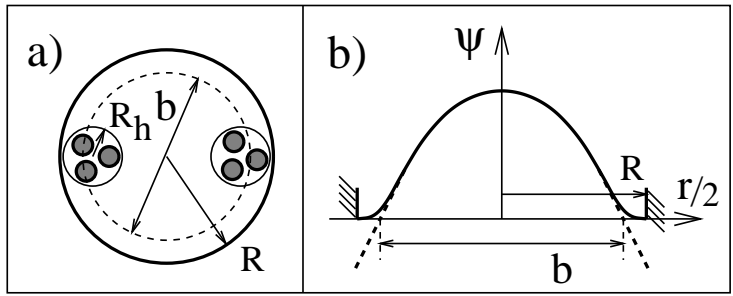}}
\medskip\medskip
\caption{
(a) A two-hadron system is confined in a spherical cavity with a radius 
$R$. Suppose this system is in an eigenstate of a definite energy 
$\e_n$.
(b) Then the wave function of the centers of its $3q$-subsystems 
(solid line) strictly vanishes at the cavity boundary. At the same 
energy $\e_n$, the wave function of unconstrained two-hadron 
motion (dashed line) vanishes when the relative hadron-hadron separation equals
$b=2R-2R_h$.}
\label{1fig}
\end{figure}    
\medskip
\be
\psi_i(b)=0\qquad\forall i=1,\ldots,n
\lb{psi=0}
\ee
then the $P$-matrix has a pole at $\e_p(b)$. As shown in Ref.~\cite{P-orig} 
its residue can be factorized:
\be
P_{ij}(\e)=\overline P_{ij}(\e)+\xi_i{~}\frac{r}{\e-\e_p(b)}{~}\xi^T_j{~}.
\lb{pole}
\ee
We have chosen the vector $\xi$ to be normalized: $\sum_{i=1}^{n}\xi_i^2=1$.

Now let us  explore the connection between the poles of the $P$-matrix and 
the quark-bag calculations. Remember that for now $b$ is taken to be 
larger than the range of the strong forces. In this case the $P$-matrix poles 
(``primitives'') $\e_p(b)$ occur at the eigenenergies of the two hadron system
with relative wave function constrained to vanish at $r=b$.  We claim that
these are just the eigenenergies\footnote{Note that we distinguish between the $P$-poles $\e_p$ and the 
eigenenergies $\e_n$ of a physical system.}, $\e_n(R)$, of the multi-quark system that
has the quantum numbers of the two-hadron system and is
confined in a spherical cavity with a  radius $R(b)$.
The radius $R$ is approximately half of $b$.
In fact, if two hadrons are placed in a hard sphere, the wave function of 
their relative motion $\psi(r)$ vanishes at  
\be
b=2R-2R_h{~},
\lb{Rbh}
\ee     
were $R_h$ plays the role of the hadron radius,
as shown in the Fig.~1.

We see that for a large value of $b$ there is one-to-one correspondence
between the $P$-matrix poles and the eigenenergies of a physical
system which is put into a hard-wall cavity. Now let us make $b$
smaller. The $P$-matrix, as defined by eq.~(\ref{PS}), will preserve
a pole structure (see eq.~(\ref{pole})) but the parameters $\e_p$, $r$, and $\xi$
will change with $b$.  If $b$ goes to $b'$ the related
$P$-pole shifts to $\e'_p$, satisfying the equation
\be
\e'_p=\e_p-\xi^T\frac{r}{\overline P(\e'_p)+k'_p{~}\cot{~}k'_p\Delta b}
{~}\xi {~},
\lb{eb}
\ee
and
\be
\Delta b = b'-b{~}.
\ee

It was noted by M.~Soldate$^{\cite{Sold}}$ 
that decreasing the cavity radius, $R$,
imposes additional constraints on the system inside and, 
therefore, causes the eigenenergies of its states $\e_n(R)$ to grow. 
In accordance with this, one can show from eq.~(\ref{eb}) that
for $b'<b$
\be
\e'_p>\e_p {~~~~~} 
\ee 
if the matrix ${\partial \overline P(\e)}/{\partial \e}$ 
is negative semidefinite, in particular if $\overline P(\e)$
is a constant.

The residue of the $P$-matrix $\xi r \xi^T$ varies with $b$ 
as
\be
\xi' r' \xi'^T=R{~}(\e'_p,\Delta b){~}\xi r \xi^T{~}R^T(\e'_p,\Delta b)
\lb{Rb}
\ee
with
\be
R{~}(\e,\Delta b)\equiv \frac{k}{\sin (k{\Delta}b)}{~}
     \frac{1}{\overline P(\e)+k{~}\cot{~}(k{\Delta}b)}{~}.
\ee
We obtained this using the $b$-independence of the 
$S$-matrix in eq.~(\ref{SP}) and the following identity:
\be
\frac{1}{A+\xi a \xi^T}=\frac{1}{A}-\frac{1}{A}{~}\xi
  \frac{a}{1+\xi^T{\displaystyle\frac{a}{A}}{~}\xi}
  \xi^T{~}\frac{1}{A}{~},
\lb{inv}
\ee
where $A$ is a nonsingular matrix, $\xi$  is a unit vector, 
and $a$ is a constant.

A small variation of $b$ in eqs.~(\ref{eb},\ref{Rb}) yields
\be
\frac{\partial\e_p}{\partial b} =-r
\lb{r}
\ee
and
\be
\frac{\partial}{\partial b}{~}(\xi r \xi^T)=
 -{~}\{\xi r\xi^T,\overline P\}\equiv -{~}\xi r\xi^T \overline P -\overline P\xi r\xi^T {~}.
\lb{R-der}
\ee 
The last equation can be converted to 
\be
\left.\overline P_{ij}(\e,b)\right|_{\e=\e_p}=-\frac{1}{2r} 
                 \frac{\partial r}{\partial b} \xi_i\xi_j
                 -\xi_i \frac{\partial \xi_j}{\partial b}
		 -\frac{\partial \xi_i}{\partial b} \xi_j+
		{~}\overline{ \overline P}_{ij}(\e_p,b)
\lb{Po}
\ee
where the matrix $\overline{ \overline P}$ is orthogonal to the vector $\xi$:
\be
\overline{ \overline P}_{ij}{~}\xi_j=\xi_j{~}\overline{ \overline P}_{ji}=0{~}.
\ee

When the cavity radius reaches the size of a few $fm$ we should 
treat the system inside as a single quark-bag rather than two hadrons,
but due to the interaction outside the bag eq.~(\ref{defP}) is no
longer valid.
Then it becomes difficult$^{\cite{Sim2}}$  to relate the quark eigenenergies
$\e_n$ to the position of the $P$ poles
$\e_p$. Nevertheless, we might expect that there is a size of the cavity 
$R_0$ when the quark-gluon system inside is already simple enough for
our theoretical tools, while $\e_n(R_0)$ and $\e_p(b_0)$ are still close 
to each other. In this case we can estimate the 
position of the $P$ poles $\e_p(b_0)$ and their orientation in the channel 
space $\xi_i(b_0)$ from quark-bag model calculations, 
and the eqs.~(\ref{r},\ref{Po}) will provide us with the other 
ingredients of the $P$-matrix.

\subsection{Calculation of P}

To specify $P$ we require the pole 
positions $\left(\e_p\right)$, the residues $\left(r^{(p)}\right)$, 
the channel coupling vectors $\left(\xi_i^{(p)}\right)$, and the nonsingular 
part $\left(\overline P_{ij}(\e)\right)$. 
We treat each of them in sequence below.

The pole positions were already considered in the previous subsection.
Let us remember that they were identified with the eigenenergies $\e_n$ of a 
quark-gluon system subject to confining boundary conditions at a sphere 
$R(b)$. 
 
In determining the vectors $\xi^{(0)}$ it is important to take account of 
flavor symmetry. For example, one may consider $SU(3)$ when describing
baryon octet scattering or $SU(2)$ for the $np$ system. For two scalar meson
scattering $SU(3)$ is badly violated and $SU(2)$ isospin symmetry is more 
appropriate.  If the flavor symmetry were exact, the mass of all hadrons 
belonging to one multiplet would be the same. The states of an interacting 
system confined by a cavity would also form flavor multiplets.
The eigenenergies of the states in one multiplet would be equal, and
the $P$-matrix would be $SU(n_{f})$ symmetric, whatever
the size of the cavity. We do not observe this in reality
because of the difference in the current quark masses. Nevertheless,
the smaller the cavity, the better the coupling vector reflects the flavor
symmetry. Let us show this in specific examples.

{\it Baryon-Baryon:}
Imagine two $\Lambda$-particles inside a {\it macroscopic} spherical cavity.
To be specific, suppose they are in the ground energy
state with $J=0$ and assume that the fusion of the $\Lambda$'s
into one $H$-dibaryon$^{\cite{H-orig}}$ is energetically forbidden, {\it
i.e.\/}
$M_H>2M_{\Lambda}$. For the macroscopic cavity the lowest eigenenergy
will be $\e_p \simeq 2M_{\Lambda}$, and the $\Lambda{-}\Lambda$ 
interaction is negligible. This $\Lambda\Lambda$ system belongs to
the symmetrized product of two $SU(3)$ baryon octets that decomposes
into the following irreducible representations:
\be
(8\otimes 8)_{\rm sym}=27\oplus 8\oplus 1{~}.
\lb{8*8}
\ee
However the $\Lambda\Lambda$ state can not be attributed to any of those 
irreducible parts, so the coupling vector $\xi_i$ in the $P$ pole 
corresponding to this state is {\it not} SU(3) symmetric.

Now we gently contract the cavity so that the system remains
in its ground state. When the cavity radius reaches
the order of 1 fm., the scale of confinement starts to 
overcome the $s$-quark mass, and $SU(3)$ symmetry gradually
emerges. The $\Lambda$'s inside split into a ``gas'' 
of 6 strongly interacting quarks. 
Due to the color-magnetic interaction, the ground state of this
system now does occur\footnote{Modulo small $SU(3)$ violation due to 
current quark masses.} at the flavor singlet$^{\cite{H-orig}}$:
\be
|H\rangle=\sqrt{\frac{1}{5}}{~}|BB\rangle+
\sqrt{\frac{4}{5}}{~}|\b 8 \cdot \b 8\rangle
\lb{Hdec}
\ee
where $\b 8\cdot\b 8$ denotes two color octet baryons coupled to an overall
singlet, and 
\begin{eqnarray}
|BB\rangle & = &\sqrt{\frac{1}{8}}{~}\left\{
|\Xi^-p\rangle-{~}|\Xi^0n\rangle+{~}|p\Xi^-\rangle-{~}|n\Xi^0\rangle
\right.\nonumber\\
& + &|\Sigma^-\Sigma^+\rangle+{~}|\Sigma^+\Sigma^-\rangle-
{~}\left.|\Sigma^0\Sigma^0\rangle+{~}|\Lambda\Lambda\rangle \right \}
\lb{BBsingl}
\end{eqnarray}
is the flavor singlet state composed of two color singlet, flavor octet 
baryons. 

Let us explore the interpretation of $\xi_i$ as the bag state orientation 
in the channel space. Consider the parameter $b$ in eq.~(\ref{defP}) 
independently for each channel. 
If at the cavity boundary the interaction is negligible,
the ``partial residue'' of the $i$-th channel will be
\be
r\xi^2_i=-\frac{\partial}{\partial b_i}\e_p
  \propto \left.\left|\frac{\partial\psi_i}{\partial r_i}
                \right|^2\right|_{r_i=b}{~}
\lb{pr}
\ee
($\psi$ is the normalized wave function of the confined system, obeying
$\left.\psi\right|_{r=b}{=}0$).
Thus $r\xi^2_i$ is associated with the ``partial pressure'' on the cavity 
walls. Eq.~(\ref{pr}) shows that for a small $b$ the residue 
$\xi_i r \xi^T_j$ of the lowest $P$-matrix pole is almost a $SU(3)$ singlet. 
As a first approximation we can take the vector $\xi$ corresponding to the
exact $SU(3)$ symmetry ({\it cf.} eq.~(\ref{BBsingl})): 
\be
\xi_i=\pm\sqrt{\frac{1}{8}}{~},{~~~~}
i=(\Xi^-p,{~}\Xi^0n,{~}p\Xi^-,{~}n\Xi^0,{~}\Sigma^-\Sigma^+,{~}\Sigma^+\Sigma^-,
{~}\Sigma^0\Sigma^0,{~}\Lambda\Lambda){~}.
\lb{xi}
\ee

{\it Meson-Meson:}
The $f_0(980)$ resonance has the quantum
numbers $I(J^{PC})=0{~}(0^{++})$ and decays strongly into
$\pi\pi$ and $\bar{K}K$. Because of the great difference between the $\pi$ and
$K$ masses, it is not
realistic to assume SU(3) symmetry even within the confinement
radius.  The SU(2) symmetric decomposition for
$f_0$ reads :
\be
|f_0\rangle=\alpha_K\sqrt{\frac{1}{4}}{~}\left\{|K^-K^+\rangle-{~}|\bar{K^0}K^0\rangle+
      |K^+K^-\rangle-{~}|K^0\bar{K^0}\rangle\right\}+
\lb{fdec}\\
 +\alpha_\pi\sqrt{\frac{1}{3}}{~}\left\{|\pi^-\pi^+\rangle+{~}|\pi^+\pi^-\rangle-
      {~}|\pi^0\pi^0\rangle\right\}+\alpha_\eta{~}|\eta\eta\rangle+{~}\alpha_{c}{~}|c\rangle 
  \nonumber
\ee
where $|c\rangle$ stands for confined channels, {\it e.g.} 
glueball, and $\Sigma|\alpha_i|^2=1$. Therefore, the $P$-matrix for $\pi\pi$
scattering has a pole around $980{~}MeV$, and its ``orientation'' in
the channel space $\xi$ is given by the normalized projection of the 
decomposition eq.~(\ref{fdec}) onto the two-particle channels $\pi\pi$, 
$\bar{K}K$, and $\eta\eta$.

Without a deeper understanding of confinement, we are only able to provide 
crude estimate for the dynamical parameters $r$ and $\overline P$.
These will serve as a guide in the next sections.  Ref.~{\cite{P-orig}} 
contains rather visual reasoning concerning the residue $r$
that we paraphrase as follows. Let us consider the ``partial pressure'' 
$p_i$ on the cavity walls due to the $i$-th flavor component of the system: 
\be
4\pi R^2p_i\equiv -{~}\frac{\partial}{\partial R_i}\e_p{~}.
\lb{pp}
\ee 
We suppose that there is a size of the cavity $R_0$ when $p_i$ can {\it either}
be calculated perturbatively in the quark-bag model {\it or}
attributed to the hadrons in the $P$-matrix approach. 
In the two-baryon example above the pressure exerted by the $\Lambda{-}
\Lambda$ subsystem is
\be
4\pi R^2 p_{\Lambda\Lambda} = 
-\frac{\partial \e_n}{\partial R}\zeta^2_{\Lambda\Lambda}
{~~~~~~~~~}\mbox{ with }{~}\zeta^2_{\Lambda\Lambda}=\frac{1}{40}{~}, 
\ee 
as calculated from the bag model (eqs.~(\ref{Hdec},\ref{BBsingl})). 
In the $P$-matrix formalism it is 
\be
4\pi R^2 p_{\Lambda\Lambda} = -{~}\frac{\partial b}{\partial R}{~}
  \frac{\partial \e_p}{\partial b_{\Lambda\Lambda}}{~}
  \xi_{\Lambda\Lambda}^2=
  \frac{\partial b}{\partial R}{~}r\xi_{\Lambda\Lambda}^2
{~~~}
\mbox{ with }{~}\xi_{\Lambda\Lambda}^2=\frac{1}{8}{~}, 
\ee 
see eqs.~(\ref{pr},\ref{xi}). That is generally we have:
\be
r\xi^2_i\simeq\frac{\partial R}{\partial b}{~}
  \frac{\partial \e_n}{\partial R}\left.\zeta^2_i\right|_{R=R_0}
{~}.
\lb{r-est}
\ee
The important result is that the ``partial residue'' 
$r\xi^2_i$ is suppressed by the factor $\zeta^2_i/\xi^2_i<1$ with 
respect to the natural scale $1/R_0^2$ .

To make the choice of $R_0$ and $b_0$, we follow the original 
paper$^{\cite{P-orig}}$. 
In Ref.\cite{P-orig} Jaffe and Low employed the MIT bag-model 
where the bag with the mass $M$ had the radius
\be
R_0\simeq 5M^{1/3}{~}GeV^{-1}{~},
\ee 
for $M$ in $GeV$. $b_0$ was obtained by matching the density of 
the free hadron-hadron wave function, vanishing as the relative
hadron separation reached $b_0$,
to the density of the free quarks inside the bag. 
In the case of two mesons this procedure yielded$^{\cite{P-orig}}$
\be
b_0\simeq 1.4{~}R_0{~}.
\lb{bR}
\ee 
 
We can say even less about the matrix $\overline P(\e)$.
Definitely, it has the poles corresponding to the other bag states.  
Eqs.~(\ref{R-der}) or (\ref{Po}) suggest that in the interstitial region
\be
\overline P_{ij}\sim \frac{1}{b_0}.
\ee

In principle, all the information about $P$-matrix 
can be rigorously obtained from calculations involving only
hadronic sizes. To this end one should solve the quark dynamics
and parameterize the hadronic wave function according to eq.(\ref{defP}).
The external interaction can be taken into account as described in 
Ref.\cite{P-orig}.  In the absence of powerful methods applicable to scales of
order 1 $fm$ we have resorted to bag model
phenomenology. 
\section{Corresponding S-matrix}

Now we turn our attention to the quantities measured in actual scattering 
experiments, such as the $S$-matrix and singularities in cross-sections.
In the Subsection 3A we express $S$ and its singularities in terms of the
$P$-matrix discussed earlier. Then we consider in detail threshold effects and
their interference with $P$-poles.

\subsection{General equations}
 
In the previous section we argued that the poles of the $P$-matrix
have fundamental significance. Taking $P$ in the form
\be
P_{ij}(\e)=\overline P_{ij}(\e)+\xi_i{~}\frac{r}{\e-\e_p}{~}\xi^T_j{~},
\lb{pole1}
\ee
one can easily reconstruct the corresponding $S$-matrix using
eq.~(\ref{SP}). In the denominator of eq.~(\ref{SP}) one has to deal with 
the inversion of a matrix like $A_{ij}+\xi_{i} a \xi^T_{j}$ and 
the identity (\ref{inv}) comes handy. After some calculations we obtain
\be
S_{ij}=\overline S_{ij}-i\chi_{i}{~}\frac{1}{\e-\e_r+
  i{~}{\displaystyle\frac{\gamma}{2}}}
       {~}\chi^T_{j}{~}.
\lb{Sgen}
\ee
In this formula $\overline S$ is the background scattering produced by 
$\overline P(\e)$:
\be
\overline S(\e)=e^{-ikb}{~}\frac{{\frac{1}{\sqrt{k}}} \overline P
  {\frac{1}{\sqrt{k}}} +i} {{\frac{1}{\sqrt{k}}} 
   \overline P {\frac{1}{\sqrt{k}}}-i}{~}e^{-ikb}{~}.
\lb{SoP}
\ee
The pole term in eq.~(\ref{Sgen}) has diverse manifestations
in cross-sections, that are discussed in the next subsection. As shorthand
for them we will make free use of the word ``resonance''.
The ``resonance'' channel couplings $\chi_i$ are 
\be
\chi_i(\e)=\sqrt{2r}{~}e^{-ik_ib}\sqrt{k_i}
 \left(\frac{1}{\overline P-ik}\right)_{ij}\xi_j{~}.
\lb{chiP}
\ee
For the energy dependent ``resonance'' position and the width in the 
denominator of eq.~(\ref{Sgen}) we have
\be
\e_r(\e)-i{~}{\displaystyle\frac{\gamma(\e)}{2}}=
  \e_p-\xi^T\frac{r}{\overline P-ik}{~}\xi{~}.
\lb{esP}
\ee
The real and imaginary parts in eq.~(\ref{esP}) can be easily separated.
To this end we write the many-channel momentum matrix $k$ as
\be
k=q+i\kappa{~},
\ee
where $q$ and $\kappa$ are real and refer to the open and closed channels
correspondingly. Recalling that for the strong interaction $\overline P$ is 
also real, we find
\be
\e_r(\e)=\e_p-{~}\xi^T
  \frac{r}{\overline P+\kappa+q
  \frac{1}{\overline P+\kappa}q}{~}\xi{~}
\lb{erP},
\ee
\be
\gamma(\e)=2r{~}\xi^T
 \frac{1}{\sqrt{1+\left(\frac{1}{\overline P+\kappa}
  q\right)^2}}{~}\frac{1}{\overline P+\kappa}{~}q{~}
 \frac{1}{\overline P+\kappa}{~}\frac{1}{\sqrt{1+\left(q
  \frac{1}{\overline P+\kappa}\right)^2}}{~}\xi{~}.
\lb{gP}
\ee
These equations are valid for a nonsingular $\overline P{+}\kappa$ and 
an arbitrary $q$.
One may write the total width in eq.~(\ref{gP}) as a sum of partial width 
$\gamma_i$ over only the open channels: 
\be
\gamma=\sum_{\stackrel{\mbox {\scriptsize open}}{\mbox
{\scriptsize channels}}}\gamma_i{~}
\lb{sumg}
\ee
with the $i$-th partial width:
\be
\gamma_i=2rq_i\left(\frac{1}{\overline P+\kappa}{~}
  \frac{1}{\sqrt{1+\left(q
  \frac{1}{\overline P+\kappa}\right)^2}}{~}\xi
 \right)_i^2{~~~~~}.
\lb{gi}
\ee

As expected, the $S$-matrix eq.~(\ref{Sgen}) is unitary,  
\be
SS^{\dagger}=1{~},
\ee
when $k$ is real.
This is ensured by the following properties of the amplitudes $\chi_i$:
\be
\overline S_{ij}\chi_j^*=\chi_i{~},
\lb{S0x}
\ee
and
\be
\chi^{\dagger}_i\chi_i=\gamma{~}.
\lb{xx}
\ee
If at some energy only the first $m<n$ channels are open, then only the 
upper-left $m\times m$ sub-matrix of $S_{ij}$  is unitary. Note that the 
unitarity of the $S$-matrix does not impose any additional restrictions on
its background part $\overline S$, except that it be unitary by itself, or 
on the $P$-matrix poles and residues. As long as $P$ is hermitian,
$S$-matrix unitarity is automatically taken care by eq.~(\ref{SP}). 

As we saw earlier, $P(\e,b_0)$ is completely determined by dynamics in the 
microscopic domain where the interaction is strong, and $P$ is not influenced
by the region in configuration space where the system is represented
by two freely moving hadrons. Thus, {\it all kinematical effects are 
absorbed in eqs.~(\ref{Sgen}-\ref{gP})}. We proceed to study them next.
\subsection{Cusp analysis}

At threshold kinematics plays a key role. The analytical structure 
of the $S$-matrix at threshold is well known and conveniently described 
via the $K$-matrix parameterization. Since the $K$ matrix can be viewed
as a special case of $P$ (see eq.~(\ref{KP})), the $P$-matrix 
formalism should provide similar results, with a different
dynamical interpretation however. In this section we outline the
analytical structure of the general equation eq.~(\ref{Sgen}) at threshold and
point out the specific features that arise from the dynamical estimates
in the  Sec.~2A~. 

Suppose the energy $\e_p$ of a $P$-pole lies at the vicinity of a threshold
in some channel, which we denote by $i=1$. The pole may correspond to
$f_0(980)$ or $a_0(980)$ at $\bar{K}K$ threshold ($990{~}MeV$);
$\Lambda(1405)$ at $\bar{K}N$ ($1430{~}MeV$); $H$-dibaryon 
($2090{-}2240{~}MeV$) at $\Lambda\Lambda$ ($2230{~}MeV$); even the
deuteron or the $pp$ virtual state at the $pn$ or $pp$ thresholds 
correspondingly.
At energies $\e_s$, satisfying the equation 
\be
\e_s=\e_r(\e_s)-i{~}{\displaystyle\frac{\gamma(\e_s)}{2}}=
     \e_p-\xi^T\frac{r}{\overline P(\e_s)-ik(\e_s)}{~}\xi {~}.
\lb{eseq}
\ee
the denominator in eq.~(\ref{Sgen}) vanishes,
{\it i.e.\/} the $S$-matrix has a pole. Note that the right hand side of
eq.~(\ref{eseq}) is a multivalued function because the momentum
\begin{figure}
\fpsxsize=2.5in
\def\fpsangle{90}
\centerline{\fpsfile{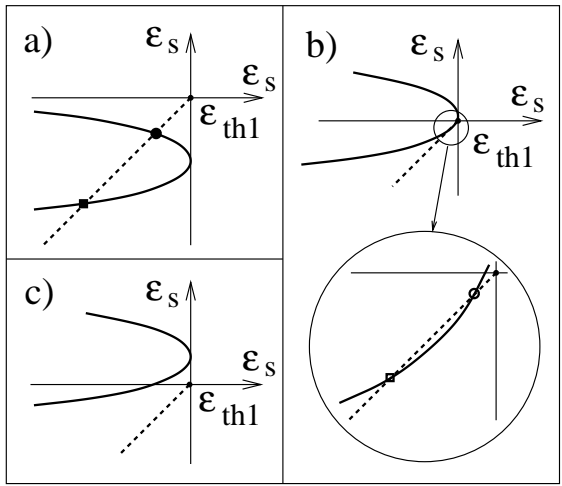}}
\medskip\medskip
\caption{
Finding $S$-matrix poles $\e_s$ as solutions of eq.~(45) for real 
energies $\e_s$ just below the lowest threshold $\et$.
The solid parabola-like curve represents the right-hand side of
eq.~(45) as a function of $\e_s$. The intersections of the curve
with the dashed diagonal line yield the solutions.
(a) There are two real energy solutions. The upper one (black circle)
corresponds to a bound state. The other one is located on the unphysical 
energy sheet.
(b) Again, the $S$-matrix has two real poles with $\e_s<\et$ 
but both of them lie on the unphysical sheet. Now the circle corresponds
to a virtual state.
(c) There are no real solutions. In this case the $S$-poles occur at complex
energies, as shown in Figure 3.}
\label{2fig}
\end{figure}
\medskip
\be
k(\e)= {\mbox{diag}}\left(\sqrt{{~}2m_1{~}(\e-\e_{\rm th1})},
\sqrt{{~}2m_2{~}(\e-\e_{\rm th2})}, \ldots,  \sqrt{{~}2m_n{~}(\e-\e_{\rm th{~}n})}
\right)\lb{ke}
\ee
has branch points at all threshold energies $\e_{\rm th{~}i}$ .
A pole in $S(\e)$ gives rise to an anomaly in the cross-section 
if only it is close enough to the physical region, where 
each $k_i$ is either real and positive (for the open channels) 
or $k_i=i\kappa_i$ with real and positive
$\kappa_i$ (for the closed channels). 
Accordingly, we are interested in the solutions of eq.~(\ref{eseq})
in which the nonsingular momenta $k_{i{\not=}1}$ are taken to be 
close to the positive real or upper imaginary semi-axis.
Having specified the branches for $k_i$ with $i{\not=}1$, 
we still in general have {\it two} sets of solutions
corresponding to different $k_1$ branches.

Let us discuss the case when at our threshold nearby the
$P$-pole, all  the other channels $i{\not=}1$ are closed. 
Consider the energies below the threshold,
\be
\e_s\leq\e_{\rm th1}<\e_{{\rm th}~i{\not=}1}{~},
\ee 
and choose the nonsingular momenta on the upper imaginary semi-axis: 
\be
k_i=i\kappa_i{~~}\mbox{with}{~~}\kappa_i>0{~},{~~}\mbox{for}{~~}i\not=1{~}.
\ee
In Fig.~2 we plot the two remaining branches 
($k_1{=}{\pm}i\kappa_1$) of the right hand side of eq.~(\ref{eseq}) 
verses $\e_s$. The solutions of eq.~(\ref{eseq}) are given by the 
intersections of this~curve 
\begin{figure}
\fpsxsize=1.5in
\def\fpsangle{90}
\centerline{\fpsfile{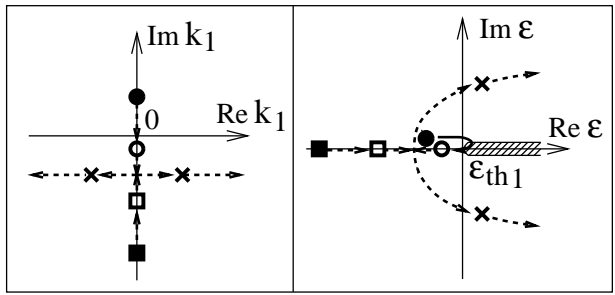}}
\medskip\medskip
\caption{
The $S$-matrix pole dynamics with the coupling strength decreasing.
The complex plane of the channel momentum $k_1$ (left) and 
the energy plane (right) are shown. Note that while the pole marked 
by the circle goes from the upper $k_1$ half-plane down to the 
lower half-plane, it moves under the cut from the physical energy 
sheet  onto the unphysical sheet.}
\label{3fig}
\end{figure}
\medskip

\noindent and the dash diagonal line. If attraction 
in the system is strong enough, we have two solutions, one on each branch 
(Fig.~2(a)). The $S$-pole with $k_1{=}i\kappa_1$ (the upper branch
in the figure) corresponds 
to a stable state. When the primitive energy $\e_p$ goes up or its residue
$r$ becomes weaker this pole moves onto the branch $k_1{=}-i\kappa_1$
(Fig.~2(b)), and the stable state turns into a virtual one. 
At last, when there is no intersection (Fig.~2(c))
the $S$-poles leave the imaginary $k_1$-axis. If one of them moves
close to the physical sheet, it will appear as a resonance in the open
scattering channel.  The evolution of the $S$-matrix poles in the $k_1$ 
and $\e$ complex planes is  shown on the Fig.~3~.

 Due to interaction with the open channels the location of a pole in 
the $S$-matrix is shifted with respect to the $P$-pole energy
$\e_p$, as given by eq.~(\ref{eseq}). For a bound state this shift,
$\e_s{-}\e_p$, is {\it negative}. In fact, if there is a bound state 
at the energy ${\cal E}$ then the wave function of the system in this state 
vanishes at infinity. From the arguments of the Sec.~2A we conclude that 
there is a $P$-matrix pole approaching ${\cal E}$ as $b{\rightarrow}\infty$.
In that section we also showed that the primitive energy $\e_p(b)$ is a 
monotonically decreasing function of $b$, therefore
\be
\e_p(b)<\e_p(\infty)={\cal E}{~}.
\lb{p0}
\ee
Of course, $S$-matrix also has a pole at
the bound state energy:
\be
\e_s={\cal E}{~}.
\lb{s0}
\ee
Combining the eqs.~(\ref{p0}) and (\ref{s0}),
\be
\e_s-\e_p<0{~}.
\ee
This result as well as an equation similar to eq.~(\ref{eseq}) were already
obtained in Ref.\cite{Kerb} for a simplified dynamical 
model$^{\cite{Sim2}}$.

Our equations~(\ref{chiP}), (\ref{esP}), or (\ref{eseq})
involve the {\it a-priori} nontrivial matrix
$(\overline P{-}ik)^{-1}$. We want to show that many of its non-diagonal 
elements at the threshold are small and may be neglected.
In fact, we mentioned in Sec.~2B that away from the other 
bag states (primitives) the characteristic scale for $\overline P$ is $1/b_0$, 
that is much less than the system mass. If for all $i{\not=}1$ 
channels
\be
\left|k_i\right|\gg\left|\overline P\right|\sim\frac{1}{b_0}{~},
\ee  
or equivalently
\be
\left|\e-\e_{\rm th{~}i}\right|\gg\frac{(1/b_0)^2}{2m_{\rm i,reduced}}
  \simeq\left\{\begin{array}{cl}10{~}MeV & \mbox{at two-baryon threshold,} \\
           30{~}MeV & \mbox{at ${\bar K}K$ threshold;}     \end{array}\right.
\lb{bigk}
\ee
then $k$ dominates the matrix $P_0{-}ik$ in every $i{\not=}1$ direction. 
Consequently, 
\be
\left(\frac{1}{P_0{-}ik}\right)_{ij}{~}<<{~}\left\{{~}
\left(\frac{1}{P_0{-}ik}\right)_{ii},{~}
\left(\frac{1}{P_0{-}ik}\right)_{1i}{~}\right\}{~}<<{~~}
\left(\frac{1}{P_0{-}ik}\right)_{11}{~},
\ee
where $i\not=j$ and $i,j\not=1$.

 With this remark in hand we can easily analyze the energy dependence of
the effective ``resonance'' position $\e_r$ (eq.(\ref{erP})), its width 
$\gamma$ (eq.(\ref{gP})), the channel couplings $\chi_i$ (eq.(\ref{chiP})), 
and hence the cross-section itself.
If at the first threshold the other $i{\not=}1$ channels are closed and 
satisfy the condition eq.(\ref{bigk}), then 
\begin{eqnarray}
\e_r & \simeq & \e_p-\sum_{i\not=1}\left(\xi_i-\xi_1\frac{\overline P_{1i}}
{\bar P_{11}}\right)\frac{r}{\kappa_i}\left(\xi_i-\frac{\overline P_{i1}}
{\overline P_{11}}{~}\xi_1\right)+
     \nonumber\\
     & + &     
     \sum_{i\not=1}\xi_1\frac{\overline P_{1i}}{\bar
 		  P_{11}}{~}\frac{r}{\kappa_i}{~}
     \frac{\overline P_{i1}}{\overline P_{11}}{~}\xi_1
    -{~}\xi_1\frac{r}{\overline P_{11}+z_1}{~}\xi_1
\label{erap}
\ee
with
\be
z_1=\left\{{\displaystyle
    \begin{array}{cl}\kappa_1,&\e<\e_{th1}{~};\\
       q_1{\displaystyle\frac{1}{\overline P_{11}}}{~}q_1,& \e>\e_{th1}{~}.
\end{array}}
\right.
\ee
(we do not consider the degenerate case $\overline P_{11}\simeq 0$).
The momentum $q_1$ or $ik_1$ in $z_1$
varies rapidly at the threshold. Together with the smallness of
$\overline P_{11}$, it may introduce dramatic energy dependence in the last 
term of eq.(\ref{erap}), as demonstrated in the Fig.~4(a)~. 
The width $\gamma$, given by eqs.(\ref{sumg}-\ref{gi}), also contains a 
rapidly changing factor $q_1$.
The experimentally {\it observed} width of the resonance or the virtual 
state is determined by both $\gamma(\e)$ {\it and} $\e_r(\e)$ . 
To show this, let us consider the resonance phase 
\be
e^{\displaystyle{~}i\varphi_r}\equiv \frac{(\e_r-\e)+i{~}
    {\displaystyle\frac{\gamma}{2}}}
  {\left|(\e_r-\e)+i{~}{\displaystyle\frac{\gamma}{2}}\right|}{~}.
\lb{phir}
\ee
\medskip
\begin{figure}[b]
\fpsxsize=3in
\def\fpsangle{90}
\centerline{\fpsfile{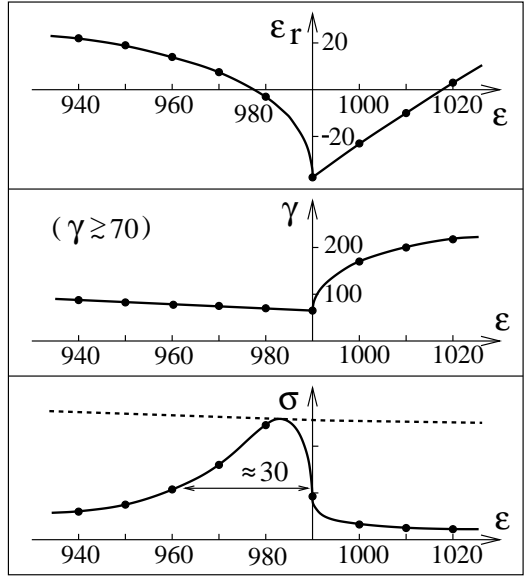}}
\medskip\medskip
\caption{
The effective resonanse position ($\e_r$), effective width 
($\gamma$), and ellastic cross section ($\sigma$) as functions of 
energy $\e$ for a two-channel model with a $P$-pole close to a 
threshold. All quantities are in $MeV$, and the threshold occurs 
at $990{~}MeV$. The half-width of the peak in the cross-section is 
signifinicantly narrowed with respect to $\gamma$.}
\label{4fig}
\end{figure}
\medskip\medskip

\noindent
Strong energy dependence of $\e_r(\e)$ may lead to 
rapid variation of $\varphi_r(\e)$ and substantial
narrowing of the observed width. An example is presented in the Fig.~4~,
where two hypothetical particles with the masses $140{~}MeV$ and $495{~}MeV$
are coupled by a $P$-pole at $1040{~}MeV$, which is $50{~}MeV$ above the 
second threshold. The couplings $\xi_1$ and $\xi_2$ are taken to be equal.
For this model the formal width $\gamma(\e)$ is no less than $70{~}MeV$
at any energy, whereas the observed half-width of the corresponding resonance
is as small as $30{~}MeV$. Of course, this simple example does not 
pretend to describe a real world and any resemblence of the Fig.~4(c) to a 
known resonance is a mere coincidence.

\section{Acknowledgments}

We are grateful to B.~Kerbikov for providing useful references and X.~Yang 
for assistance with numerical work.

\newpage 

\end{document}